\documentclass[a4paper,usenatbib,longnamesfirst]{mnras}

\usepackage{newtxtext,newtxmath}
\usepackage[T1]{fontenc}
\usepackage{ae,aecompl}

\usepackage{amsmath}
\usepackage{amsfonts}
\usepackage{amssymb}
\usepackage{graphicx}
\usepackage{color}
\usepackage{hyperref}
\usepackage{booktabs}
\usepackage{hyperref}
\usepackage{cleveref}
\usepackage{color}
\usepackage{multirow}
\usepackage{lipsum}

\Crefname{equation}{Eq.}{Eqs.}
\Crefname{figure}{Fig.}{Figs.}
\Crefname{section}{Sec.}{Secs.}

\usepackage{etoolbox}
\makeatletter
\appto{\appendix}{%
  \@ifstar{\def\thesection{\unskip}\def\theequation@prefix{A.}}%
          {\def\thesection{\Alph {section}}}%
}
\makeatother

\title[]{Addressing the circularity problem in the $E_\text{p}-E_\text{iso}$ correlation of Gamma-Ray Bursts}

\author[]{
Lorenzo Amati,$^{1}$\thanks{lorenzo.amati@inaf.it}
Rocco D'Agostino,$^{2}$\thanks{rocco.dagostino@roma2.infn.it}
Orlando Luongo,$^{3,4}$\thanks{orlando.luongo@lnf.infn.it}\newauthor
Marco Muccino,$^{3}$\thanks{marco.muccino@lnf.infn.it}
Maria Tantalo$^{5}$\thanks{maria.tantalo@lnf.infn.it}
\\
$^{1}$INAF, Istituto di Astrofisica Spaziale e Fisica Cosmica, Bologna, Via Gobetti 101, I-40129 Bologna, Italy.\\
$^{2}$Sezione INFN, Universit\`{a} di Roma ``Tor Vergata'', Via della Ricerca Scientifica 1, I-00133, Roma, Italy.\\
$^{3}$Istituto Nazionale di Fisica Nucleare, Laboratori Nazionali di Frascati, 00044 Frascati, Italy.\\
$^{4}$NNLOT, Al-Farabi Kazakh National University, Al-Farabi av. 71, 050040 Almaty, Kazakhstan.\\
$^{5}$Dipartimento di Fisica, Universit\`{a} di Roma ``Tor Vergata'', Via della Ricerca Scientifica 1, I-00133, Roma, Italy.
}

\date{Accepted XXX. Received YYY; in original form ZZZ}

\pubyear{2019}

\begin{document}
\label{firstpage}
\pagerange{\pageref{firstpage}--\pageref{lastpage}}
\maketitle

\begin{abstract}
We here propose a new model-independent technique to overcome the circularity problem affecting the use of Gamma-Ray Bursts (GRBs) as distance indicators through the use of $E_{\rm p}$--$E_{\rm iso}$ correlation. We calibrate the $E_{\rm p}$--$E_{\rm iso}$ correlation and find the GRB distance moduli that can be used to constrain dark energy models. We use observational Hubble data to approximate the cosmic evolution through B\'ezier parametric curve obtained through the linear combination of Bernstein basis polynomials. In so doing, we build up a new data set consisting of 193 GRB distance moduli. We combine this sample with the supernova JLA data set to test the standard $\Lambda$CDM model and its $w$CDM extension. We place observational constraints on the cosmological parameters through Markov Chain Monte Carlo numerical technique. Moreover, we compare the theoretical scenarios by performing the AIC and DIC statistics. For the $\Lambda$CDM model we find $\Omega_m=0.397^{+0.040}_{-0.039}$ at the $2\sigma$ level, while for the $w$CDM model we obtain $\Omega_m=0.34^{+0.13}_{-0.15}$ and $w=-0.86^{+0.36}_{-0.38}$ at the $2\sigma$ level. Our analysis suggests that $\Lambda$CDM model is statistically favoured over the $w$CDM scenario. No evidence for extension of the $\Lambda$CDM model is found.
\end{abstract}




\section{Introduction}
\label{intro}

The cosmic speed up is today a consolidate  experimental evidence confirmed by several probes \citep{Haridasu17}. Particularly, type Ia Supernovae (SNe Ia) have been employed  as standard candles \citep{Phillips1993} to check the onset of cosmic acceleration \citep{1998Natur.391...51P,Perlmutter1999,Riess1998,Schmidt1998}. Their importance lies in the fact that they may open a window into the nature of the constituents pushing up the universe to accelerate. Even though SNe Ia are considered among the most reliable standard candles, they are detectable at most at redshifts $z\simeq 2$ \citep{Rodney2015}. Thus, at intermediate redshifts the standard cosmological model, dubbed the $\Lambda$CDM paradigm, cannot be tested with SNe Ia alone.
Consequently, higher redshift distance indicators, such as Baryon Acoustic Oscillations (BAO) \citep{Percival10,Aubourg15,Lukovic16}, have been used to alleviate degeneracy among the $\Lambda$CDM paradigm and dark energy scenarios.
In these respects, a relevant example if offered by Gamma-Ray Bursts (GRBs), which represent the most powerful cosmic explosions detectable up to $z=9.4$ \citep{Salvaterra2009,Tanvir2009,Cucchiara2011}.
Attempts to use GRBs as cosmic rulers led cosmologists to get several correlations between GRB photometric and spectroscopic properties \citep{Amati2002,Ghirlanda04,Amati2008,Schaefer2007,CapozzielloIzzo2008,Dainotti2008,Bernardini2012,AmatiDellaValle2013,Wei2014,Izzo2015,Demianski17a,Demianski17b}.
The most investigated correlations involve the rest-frame spectral peak energy $E_{\rm p}$, i.e. the rest-frame photon energy at which the $\nu$F$_\nu$ spectrum of the GRB peaks, and the bolometric isotropic-equivalent radiated energy $E_{\rm iso}$, or peak luminosity $L_{\rm p}$ \citep{Amati2002, Yonetoku2004, Amati2008, AmatiDellaValle2013, Demianski17a,Demianski17b}.
However, the use of GRBs for cosmology is still affected by some uncertainties due to selection and instrumental effects and the so-called \textit{circularity problem} \citep[see, e.g.,][]{Kodama2008}. The former issue has been investigated in several studies, with the general, even though still debated, conclusion that these effects should be minor \citep[see, e.g.,][] {Amati2006, Ghirlanda06, Nava2012, AmatiDellaValle2013, Demianski17a}.
The circularity problem arises from the fact that, given the lack of very low-redshift GRBs, the correlations between radiated energy or luminosity and the spectral properties are established assuming a background cosmology. For example, calibrating GRBs through the standard $\Lambda$CDM model, the estimate of cosmological parameters of any dark energy framework inevitably returns an overall agreement with the concordance model.

In this paper, we propose a new model-independent calibration of the $E_{\rm p}$--$E_{\rm iso}$ correlation \citep[the \textit{Amati relation} see e.g.,][]{Amati2008,AmatiDellaValle2013}.
We take the most recent values of observational Hubble Data (OHD), consisting of $31$ points of Hubble rates got at different redshifts \citep[see][and references therein]{2018MNRAS.476.3924C}. These data have been obtained through the differential age method applied to pairs of nearby galaxies, providing model-independent measurements \citep{2002ApJ...573...37J}.
We follow the strategy to fit OHD data using a B\'ezier parametric curve obtained through the linear combination of Bernstein basis polynomials. This treatment is a refined approximated method and reproduces Hubble's rate at arbitrary redshifts without assuming an \emph{a priori} cosmological model.
We thus use it to calibrate the $E_{\rm iso}$ values by means of a data set made of $193$ GRBs (with firmly measured redshift and spectral parameters taken from \citealt{Demianski17a} and references therein), and compute the corresponding GRB distance moduli $\mu_{\rm GRB}$ and the $1\sigma$ error bars, depending upon the uncertainties on GRB observables.
Detailed discussions of possible biases and selection effects can be found, e.g. in \citet{AmatiDellaValle2013}, \citet{Demianski17a} and \citet{Dainotti18}. From the above model-independent analysis over OHD data, we obtain $H_0=67.74$~km~s$^{-1}$~Mpc$^{-1}$, compatible with the current estimates by the \citet{Planck18} and \citet{2018ApJ...861..126R}.

As a pure example of fitting procedure, we analyze our data by means of Markov Chain Monte Carlo (MCMC) technique and compare them with the standard cosmological paradigm and its simplest extension, namely the $w$CDM model. We discuss the limits over our technique in view of the most recent bias and problems related to SN Ia and GRBs. Afterwards, using the above value of $H_0$ got from our parametric fit analysis, we show that our results are in tension with the concordance paradigm \citep{Planck18} at $\geq 3\sigma$. However, we propose that such results may be affected by systematics and how these limits may be reconsidered in view of future developments. Finally, we compare the statistical performance of the cosmological models through the Bayesian selection criteria.

The paper is divided into four sections. After this Introduction, in Sec.~\ref{sec:calibration} we describe the main features of our treatment, using OHD data surveys over the Amati relation. In Sec.~\ref{res}, we discuss our numerical outcomes concerning the use of our new data set. We thus get constraints over the free parameters of the $\Lambda$CDM and $w$CDM models. In Sec.~\ref{conclusions}, we draw conclusions and identify the perspectives of our work.

\section{Model-independent calibration of the Amati relation}\label{amatical}
\label{sec:calibration}

Calibrating the Amati relation represents a challenge due to the problem of circularity \citep[see, e.g.,][]{Ghirlanda04,Ghirlanda06,Kodama2008,AmatiDellaValle2013}.
In fact, in the $E_{\rm p}$--$E_{\rm iso}$ correlation, the cosmological parameters $\Omega_i$ and the Hubble constant $H_0$ enter in the $E_{\rm iso}$ definition through the luminosity distance $d_{\rm L}$, i.e.,
$E_{\rm iso}\left(z,H_0,\Omega_i\right)\equiv 4\pi d_{\rm L}^2\left(z,H_0,\Omega_i\right) S_{\rm bolo}/(1+z)$,
where $S_{\rm bolo}$ is the observed bolometric GRB fluence and the factor $(1+z)^{-1}$ transforms the observed GRB duration into the source cosmological rest-frame one.
The most quoted approach to the calibration of the Amati relation makes use of the SN Ia Hubble diagram, directly inferred from the observations, and interpolate it to higher redshift by using GRBs \citep[see, e.g.,][]{Kodama2008,2008ApJ...685..354L,Demianski17a,Demianski17b}. However, this method biases the GRB Hubble diagram by introducing the systematics of the SNe Ia.

\begin{figure}
\centering
\includegraphics[width=\hsize,clip]{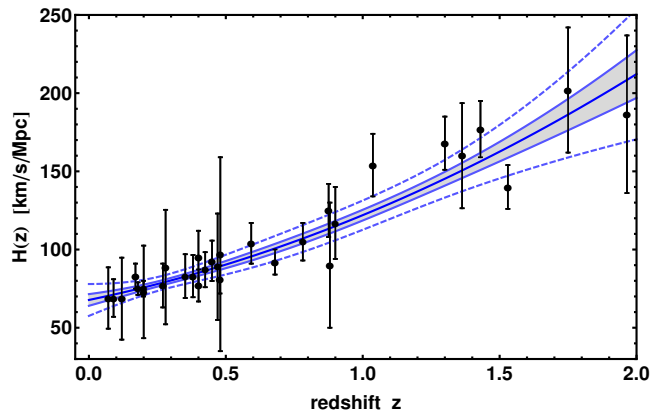}
\caption{OHD data ($31$ black points with vertical error bars), their best fit function (solid thick blue curve) and its $1\sigma$ (blue curves and light blue shaded are) and $3\sigma$ (blue dashed curves) confidence regions.}
\label{fig:1}
\end{figure}

Here, we propose an alternative calibration which makes use of the \textit{differential age method} based on spectroscopic measurements of the age difference $\Delta t$ and redshift difference $\Delta z$ of couples of passively evolving galaxies that formed at the same time \citep{2002ApJ...573...37J}. This method implies that $\Delta z/\Delta t\equiv dz/dt$ and hence the Hubble function can be computed in a cosmology-independent way as $H(z)=-(1+z)^{-1}\Delta z/\Delta t$.
The updated sample of $31$ OHD \citep[see][]{2018MNRAS.476.3924C} is shown in Fig.~\ref{fig:1}.
To avoid the circularity problem, we approximate the OHD data by employing a B\'ezier parametric curve\footnote{B\'ezier curves are easy to use in computation, are stable at the lower degrees of control points and can be rotated and translated by performing the operations on the points.} of degree $n$
\begin{equation}
\label{bezier}
H_n(z)=\sum_{d=0}^{n} \beta_d h_n^d(z)\quad,\quad h_n^d(z)\equiv \frac{n!(z/z_{\rm m})^d}{d!(n-d)!} \left(1-\frac{z}{z_{\rm m}}\right)^{n-d}\,,
\end{equation}
where $\beta_d$ are coefficients of the linear combination of Bernstein basis polynomials $h_n^d(z)$, positive in the range $0\leq z/z_{\rm m}\leq1$, where $z_{\rm max}$ is the maximum $z$ of the OHD dataset.
For $d=0$ and $z=0$, we easily identify $\beta_0\equiv H_0$.
Besides the simple cases with $n=0$ and $n=1$ leading to a constant value and a linear growth with $z$ of $H(z)$, respectively, the only case providing a monotonic growing function over the limited range in redshift of the OHD data is $n=2$; higher values lead to oscillatory behaviors of the approximating function. Therefore, in the following we use $n=2$ in fitting the OHD data. 
The best fit with its $1\sigma$ and $3\sigma$ confidence regions are shown in Fig.~\ref{fig:1}.
The best-fit parameters are $H_0=67.76\pm3.68$, $\beta_1=103.34\pm11.14$, and $\beta_2=208.45\pm14.29$ (all in units of km~s$^{-1}$~Mpc$^{-1}$).
The value of $H_0$ so obtained is compatible with the current estimate of the Planck Collaboration \citep{Planck18} and in agreement at the $1.49\sigma$ level with the value measured by \citet{2018ApJ...861..126R}.
\begin{figure}
\centering
\includegraphics[width=\hsize,clip]{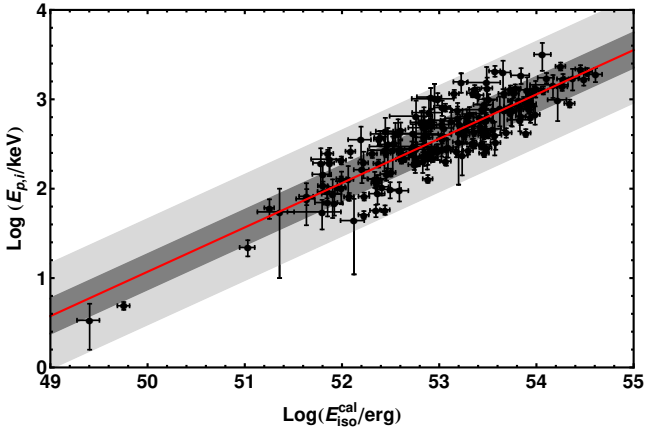}
\caption{GRB calibrated distribution in the $E_{\rm p}$--$E_{\rm iso}^{\rm cal}$ plane (black datapoints), the best-fitting function (red solid line) and the $1\sigma_{\rm ex}$ and $3\sigma_{\rm ex}$ limits (dark-gray and light-gray shaded regions, respectively).}
\label{fig:2}
\end{figure}

Once the function $H_2(z)$ is extrapolated to redshift $z>z_{\rm m}$, the luminosity distance is \citep[see, e.g.,][]{Goobar1995}
\begin{equation}
\label{dlHz}
d_{\rm L}\left(\Omega_k,z\right)=\frac{c}{H_0}\dfrac{(1+z)}{\sqrt{|\Omega_k|}}S_k\left[\sqrt{|\Omega_k|}\int_0^z\dfrac{H_0 dz'}{H_2(z')}\right] ,
\end{equation}
where $\Omega_k$ is the curvature parameter, and $S_k(x)=\sinh(x)$ for $\Omega_k>0$, $S_k(x)=x$ for $\Omega_k = 0$, and $S_k(x)=\sin(x)$ for $\Omega_k<0$.
We note that $d_{\rm L}$ in Eq.~\eqref{dlHz} is not completely independent from cosmological scenarios, since it depends upon $\Omega_k$.
However, supported by the most recent Planck results \citep{Planck18}, which find $\Omega_k=0.001\pm0.002$, we can safely assume $\Omega_k=0$.
In so doing, the dependency upon $\Omega_k$ identically vanishes and Eq.~\eqref{dlHz} becomes cosmology-independent:
\begin{equation}
\label{dlHz2}
d_{\rm cal}(z)=c(1+z)\int_0^z\dfrac{dz'}{H_2(z')}\,.
\end{equation}
We are now in the position to use $d_{\rm cal}(z)$ to calibrate the isotropic energy $E_{\rm iso}^{\rm cal}$ for each GRB fulfilling the Amati relation\footnote{
Recent works claim that our universe has non-zero curvature and that $\Omega_k$ represents at most the $2$\% of the total universe energy density (see, e.g., \citealt{2018ApJ...864...80O}, and references therein). Relaxing the assumption $\Omega_k=0$, since its value is still very small, the circularity problem is not completely healed, but it is only restricted to the value of $\Omega_k$, since $H(z)$ can be still approximated by the function $H_2(z)$.}
\begin{equation}
\label{Eisocal}
E_{\rm iso}^{\rm cal}(z)\equiv 4\pi d_{\rm cal}^2(z) S_{\rm bolo}(1+z)^{-1}\,,
\end{equation}
where the respective errors $\sigma E_{\rm iso}^{\rm cal}$ depend upon the GRB systematics on the observables and the fitting procedure (see confidence regions in Fig.~\ref{fig:1}).
The corresponding $E_{\rm p}$--$E_{\rm iso}^{\rm cal}$ distribution is displayed in Fig.~\ref{fig:2}.
Following the method by \citet{Dago2005}, we fit the calibrated Amati relation by using a linear fit $\log(E_{\rm p}/1{\rm keV})=q+m[\log(E_{\rm iso}^{\rm cal}/{\rm erg})-52]$.
We find the best-fit parameters $q=2.06\pm0.03$, $m=0.50\pm0.02$, and the extra-scatter $\sigma_{\rm ex}=0.20\pm0.01$~dex (see Fig.~\ref{fig:2}). The corresponding Spearman's rank correlation coefficient is $\rho_s=0.84$ and the p-value from the two-sided Student's $t$-distribution is $p=2.42\times10^{-36}$.

\begin{figure}
\centering
\includegraphics[width=\hsize,clip]{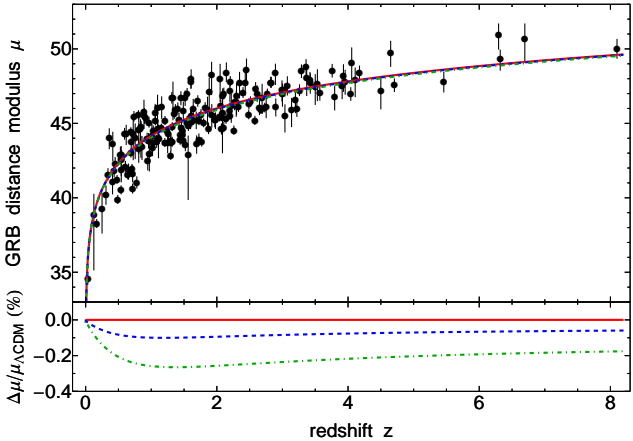}
\caption{\textit{Upper plot}: GRB distance moduli $\mu_{\rm GRB}$ distribution compared to the $\Lambda$CDM model $\mu_{\Lambda{\rm CDM}}$ with $H_0=67.36$~km~s$^{-1}$~Mpc$^{-1}$, $\Omega_m=0.3166$ and $\Omega_\Lambda=0.6847$ as in \citet{Planck18} (solid red curve), and two $w$CDM models with the above $\Lambda$CDM parameters and $w=-0.90$ (dashed blue curve) and $w=-0.75$ (dot-dashed green curve). \textit{Lower plot}: the deviations of the above three models $\mu_X$ from $\mu_{\Lambda{\rm CDM}}$ computed as $(\mu_X-\mu_{\Lambda{\rm CDM}})/\mu_{\Lambda{\rm CDM}}$ (curves retain the same meaning as before).
}
\label{fig:3}
\end{figure}

We can then compute the GRB distance moduli from the standard definition $\mu_{\rm GRB}=25+5\log(d_{\rm cal}/{\rm Mpc})$. Using the fit of the calibrated Amati relation, we obtain
\begin{equation}
\label{muGRB}
\mu_{\rm GRB}=25+\frac{5}{2}\left[\frac{\log E_{\rm p}-q}{m}-\log\left(\frac{4\pi S_{\rm bolo}}{1+z}\right)+52\right]\,,
\end{equation}
where now $S_{\rm bolo}$ has been normalized to erg~Mpc$^{-2}$ to obtain $d_{\rm cal}$ in the desiderd units of Mpc. The attached errors on $\mu_{\rm GRB}$ take into account the GRB systematics and the statistical errors on $q$, $m$ and $\sigma_{\rm ex}$.
The distribution of $\mu_{\rm GRB}$ with $z$ is shown in Fig.~\ref{fig:3}


\begin{table}
\small
\begin{center}
\caption{Priors used for parameters estimate in the MCMC analysis.}
 \label{tab:priors}
\renewcommand{\arraystretch}{1.}
\begin{tabular}{cccccc}
\hline
\hline
$w$ & $\Omega_m$  & $M$ & $\Delta_M$  &  $\alpha$ & $\beta$ \\
\hline
$(-0.5,-1.5)$  & $(0,1)$ & $(-20,-18)$ & $(-1,1)$ & $(0,1)$ & $(0,5)$ \\
\hline\hline
\end{tabular}
\end{center}
\end{table}

\begin{figure}
\centering
\includegraphics[width=\hsize,clip]{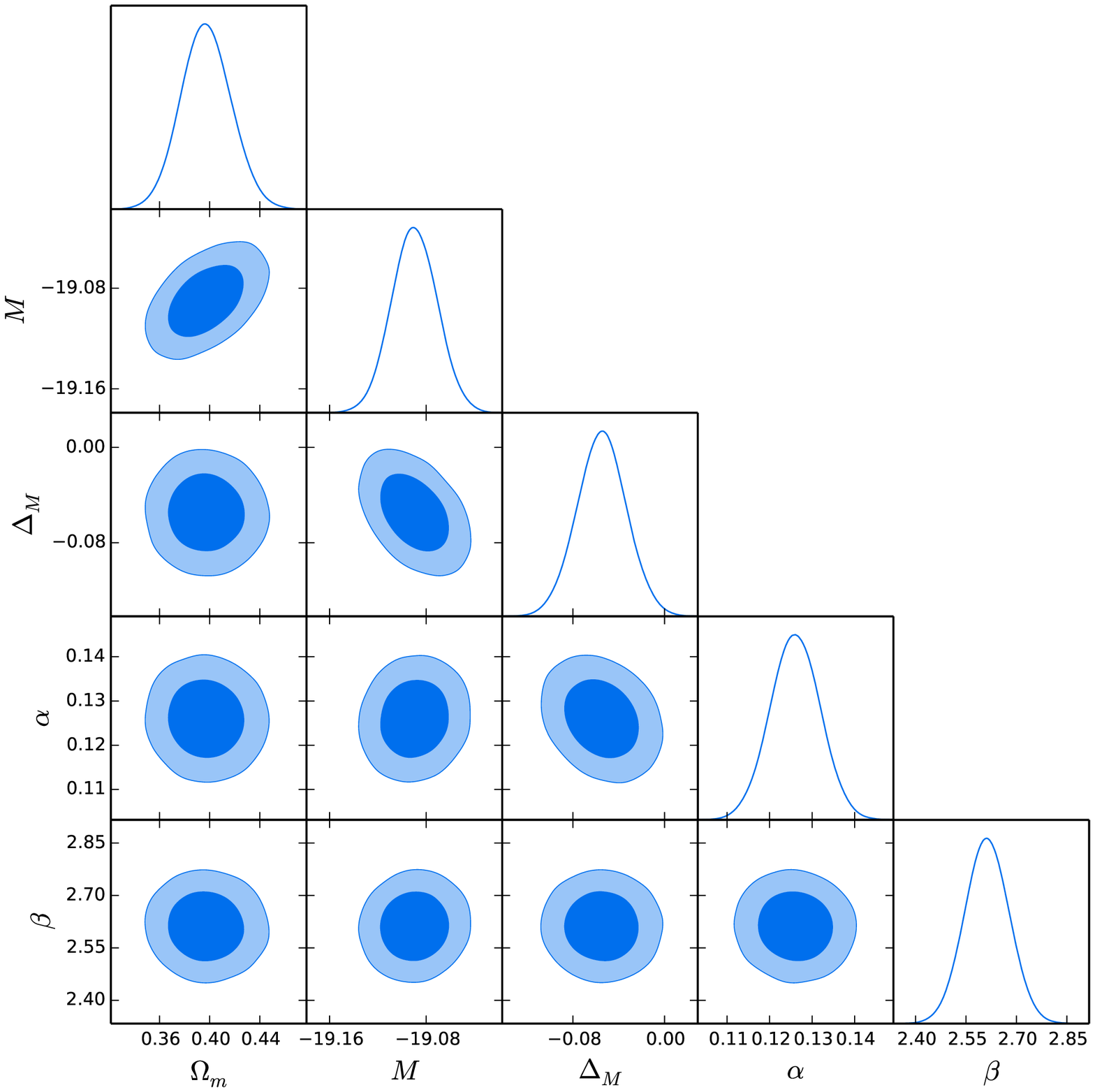}
\includegraphics[width=\hsize,clip]{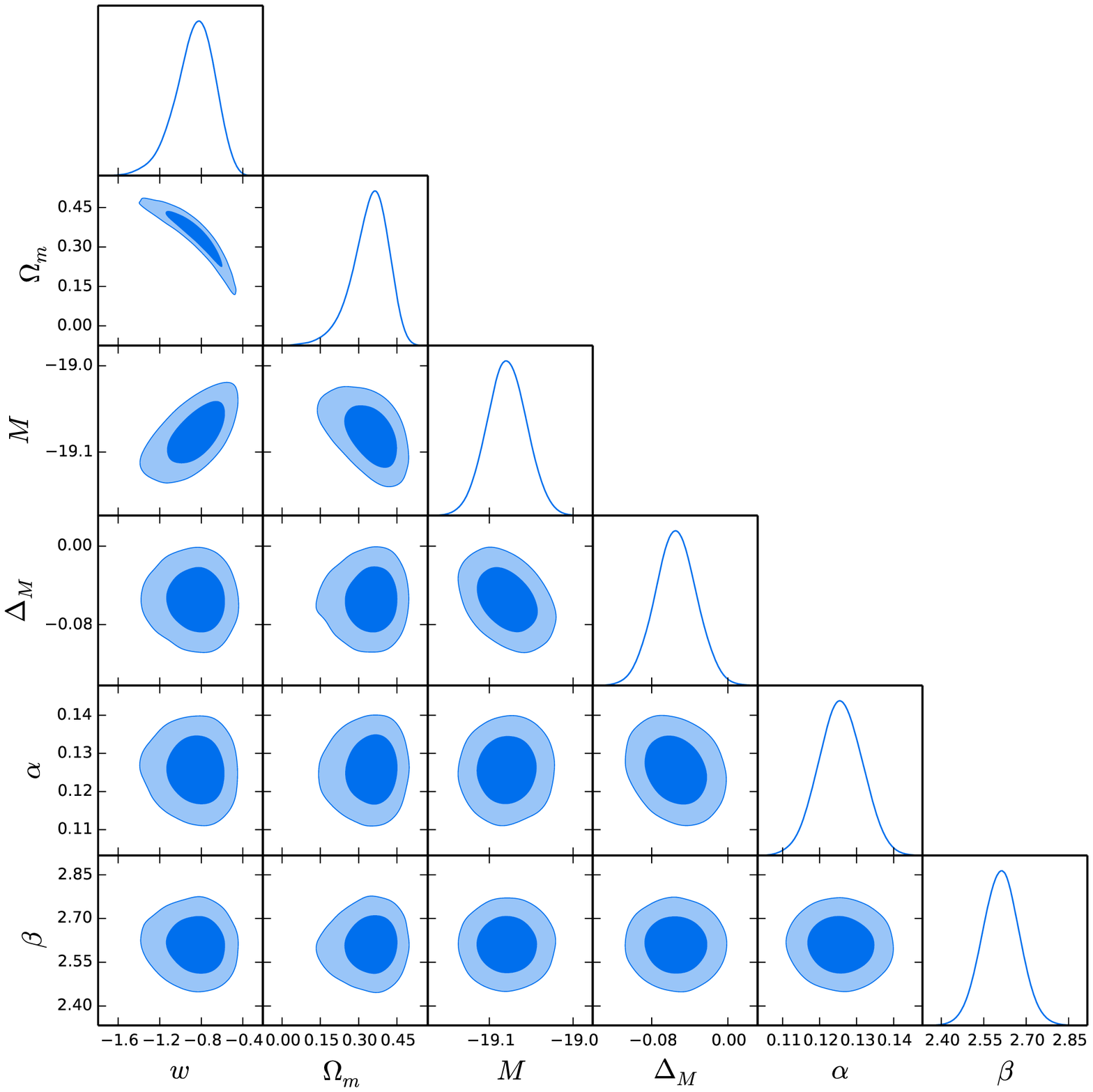}
\caption{Marginalized 1$\sigma$ and 2$\sigma$ contours, and posterior distributions from the MCMC analysis of SN+GRB data for the $\Lambda$CDM model (top) and for the $w$CDM model (bottom).}
\label{fig:contours_LCDM}
\end{figure}

We note that the statistical method adopted for the GRBs calibration may be in principle used also for the analysis of the SN data. This would in fact reduce the propagation errors when the combined fit of both data sets is performed, making the joint sample homogeneous for the cosmological studies. It will be interesting to analyze the impact of such a procedure in a forthcoming study, where the Philips relation of SN will be calibrated in the way we attempted with GRBs prior to performing the cosmological fit.

\section{Numerical results}
\label{res}

We here use our sample of GRBs to test cosmological models. In particular, we assume standard barotropic equation of state (EoS). Thus, for each fluid the pressure $P_i$ is a one-to-one function of the density $\rho_i$: $P_i=w_i\rho_i$. As a consequence of Bianchi's identity, one gets $\dot{\rho_i}+3H\rho_i(1+w_i)=0$ for each species entering the Einstein equations. Following the standard recipe, we here  consider pressureless matter with negligible radiation and define current total density as $\Omega_i=\rho_i/\rho_c$, with $\rho_c\equiv 8\pi G/(3H_0^2)$ is the critical density,
 one can reformulate the Hubble evolution as:
\begin{equation}\label{acca}
H(z)= H_0\sqrt{\Omega_{m}(1+z)^3+\Omega_{DE}(1+z)^{3(1+w)}}\ .
\end{equation}
In the above relation, dark energy takes a net density given by $\Omega_{DE}=1-\Omega_m$ to guarantee that $H(z=0)=H_0$, and $w$ is the dark energy EoS parameter.
In particular, Eq.~\eqref{acca} reduces to the $\Lambda$CDM model as $w=-1$, whereas to the $w$CDM model when $w$ is free to vary.
\begin{table*}
\small
\begin{center}
\setlength{\tabcolsep}{1em}
\renewcommand{\arraystretch}{1.2}
\caption{95\% confidence level results of the MCMC analysis for the SN+GRB data. The AIC and DIC differences are intended with respect to the $\Lambda$CDM model.}
\begin{tabular}{c c c c c c c c c}
\hline
\hline
Model & $w$ & $\Omega_{m}$ &  $M$  & $\Delta_M$ & $\alpha$ & $\beta$ & $\Delta$AIC & $\Delta$DIC\\
\hline
$\Lambda$CDM &  -1& $0.397^{+0.040}_{-0.039}$   & $-19.090^{+0.037}_{-0.037}$  & $-0.055^{+0.043}_{-0.043}$ & $0.126^{+0.011}_{-0.012}$ & $2.61^{+0.13}_{-0.13}$ & 0 & 0 \\
$w$CDM & $-0.86^{+0.36}_{-0.38}$ & $0.34^{+0.13}_{-0.15}$ & $-19.079^{+0.046}_{-0.046} $ & $-0.055^{+0.042}_{-0.042}  $ & $0.126^{+0.011}_{-0.012} $ & $2.61^{+0.13}_{-0.13} $ & $1.44$ & $1.24$ \\
\hline
\hline
\end{tabular}
 \label{tab:results}
\end{center}
\end{table*}
The distance modulus is given by $\mu_{th}(z)=25+5\log[d_L(z)/\text{Mpc}]$, where $d_L(z)$ is given by Eq.~\eqref{dlHz} with $\Omega_k=0$. Thus, the likelihood function of the GRB data can be written as
\begin{equation}
\label{Likelihood}
\mathcal{L}_{\rm GRB}=\prod_{i=1}^{N_{\rm GRB}}\frac{1}{\sqrt{2\pi}\sigma_{\mu_{\text{GRB},i}}}
\exp\left[-\dfrac{1}{2}\left(\dfrac{\mu_{th}(z_i)-\mu_{\text{GRB},i}}{\sigma_{\mu_{\text{GRB}},i}}\right)^2\right]\,,
\end{equation}
where $N_\text{GRB}=193$ is the number of GRB data points.
To obtain more robust observational bounds on cosmological parameters, we consider a complete Hubble diagram by complementing the GRB measurements with the SN JLA sample \citep{Betoule14}. The latter consists of 740 SN Ia in the redshift range $0.01<z<1.3$. The distance modulus of each SN is parameterized as
\begin{equation}
\mu_\text{SN}=m_B-M_B+\alpha X_1-\beta C\ ,
\end{equation}
where $m_B$ is the $B$-band apparent magnitude, while $C$ and $X_1$ are the colour and the stretch factor of the light curve, respectively; $M_B$ is the absolute magnitude defined as
\begin{equation}
M_B=
\begin{cases}
 M & \text{if}\ M_{host}<10^{10}M_{Sun}\ ,\\
  M+\Delta_M & \text{otherwise}\ ,
 \end{cases}
\end{equation}
where $M_{host}$ is the host stellar mass, and $M$, $\alpha$ and $\beta$ are nuisance parameters which enter the fits  along with cosmological parameters.
The likelihood function of the SN data is given as
\begin{equation}
\mathcal{L}_{SN}=\dfrac{1}{{|2\pi \mathcal{M}|}^{1/2}}\exp\left[-\dfrac{1}{2}\left(\mu_{th}-\mu_\text{SN}\right)^\text{T} \mathcal{M}^{-1}\left(\mu_{th}-\mu_\text{SN}\right)\right],
\end{equation}
where $\mathcal{M}$ is the $3N_\text{SN}\times N_\text{SN}=2220 \times 2200$ covariance matrix with the statistical and systematic uncertainties on the light-curve parameters given in \cite{Betoule14}.

We thus perform a MCMC integration on the combined likelihood function $\mathcal{L}=\mathcal{L}_\text{SN} \mathcal{L}_\text{GRB}$ by means of the Metropolis-Hastings algorithm implemented through the Monte Python code \citep{Audren13}.
In the numerical procedure, we assume uniform priors on the fitting parameters (see Table~\ref{tab:priors}) and we take $H_0$ as the best-fit value obtained from the model-independent analysis of the OHD data: $H_0=67.74$ km/s/Mpc.
We summarize the results for the $\Lambda$CDM and $w$CDM models in Table~\ref{tab:results}. We show the marginalized $1\sigma$ and $2\sigma$ confidence contours in Fig.~\ref{fig:contours_LCDM}.
One immediately sees that $\Omega_m$ in the $\Lambda$CDM model is unusually high compared to previous findings which use SNe Ia and other surveys different from GRBs. In fact, our result is in tension with Planck's predictions \citep{Planck18} at $\geq 3\sigma$.
However, our outcome is well consistent within $1\sigma$ with previous analyses which made use of GRBs (see. e.g. \citealt{AmatiDellaValle2013} for a review, and \citealt{Izzo2015}, \citealt{Haridasu17} and \citealt{Demianski17a,Demianski17b} for recent results).
In addition, the tension is reduced as one considers the $w$CDM model, enabling $w$ to vary. This does not indicate that $w$CDM is favoured with respect to the standard cosmological model. In fact, we immediately notice that $w$ is consistent within $1\sigma$ with the $\Lambda$CDM case, i.e. $w=-1$.

We note that the numerical approach using the Metropolis-Hasting algorithm may suffer from some issues related to random walk behaviour. In the case of highly correlated statistical models, the use of more robust integration methods could alleviate many of those issues. Alternative approaches for the multi-level structure of the proper Bayesian model are left for a future study.

\subsection{Statistical performances with GRBs}

To test the statistical performance of the models under study, we apply the AIC criterion (Akaike 1974):
\begin{equation*}
\text{AIC}\equiv\ 2p -2\ln \mathcal{L}_{max},\quad
\end{equation*}
where $p$ is the number of free parameters in the model and $\mathcal{L}_{max}$ is the maximum probability function calculated at the best-fit point. The best model is the one that minimizes the AIC value.
We also use the DIC criterion \citep{Kunz2006} defined as
\begin{equation*}
\text{DIC}\equiv2p_{eff}-2\ln \mathcal{L}_{max}\ ,
\end{equation*}
where $p_{eff}=\langle-2\ln\mathcal{L}\rangle+2\ln\mathcal{L}_{max}$ is the number of parameters that a dataset can effectively constrain.
Here, the brackets indicate the average over the posterior distribution.
Unlike the AIC and BIC criteria, the DIC statistics does not penalize for the total number of free parameters of the model, but only for those which are constrained by the data \citep{Liddle2007}. We thus computed the  differences with respect to the reference $\Lambda$CDM flat scenario. Both the AIC and DIC results indicate that the $\Lambda$CDM model is only slightly favoured with respect to the $w$CDM model (see Table~\ref{tab:results}).

\section{Final outlooks and perspectives}
\label{conclusions}

In this work, we faced out the circularity problem in using GRBs as distance indicators. To do so, we employed the $E_{\rm p}-E_\text{iso}$ (``Amati") correlation and we proposed a new technique to build $d_L$ in a model-independent way, using the OHD measurements. In particular, we considered the OHD data points and we approximated the Hubble function by means of a B\'ezier parametric curve obtained from the linear combinations of Bernstein's polynomials.
Assuming vanishing spatial curvature as suggested by Planck's results, we were able to calibrate the Amati relation in a model-independent way. We thus obtained a new sample of distance moduli for 193 different GRBs (see Table~\ref{tab:grbmuz}).

We then used the new data sample to constrain two different cosmological scenarios: the concordance $\Lambda$CDM model, and the $w$CDM model, with the dark energy EoS parameter is free to vary.
Hence, we performed a Monte Carlo integration through the Metropolis-Hastings algorithm on the joint likelihood function obtained by combining the GRB measurements with the SNe JLA data set. In our numerical analysis, we fixed $H_0$ to the best-fit value obtained from the model-independent analysis over OHD data, i.e. $H_0=67.74$~km~s$^{-1}$~Mpc$^{-1}$.
Our results for $\Omega_m$ and $w$ agree with previous findings making use of GRBs and our treatment candidates as a severe alternative to calibrate the Amati relation in a model-independent form. Finally, we employed the AIC and BIC selection criteria to compare the statistical performance of the investigated models. We found that the $\Lambda$CDM model is preferred with respect to the minimal $w$CDM extension.
Although a pure $\Lambda$CDM model is statistically favoured, we note that
the values of $\Omega_m$ and $w$ for the $w$CDM model are remarkably in
agreement with those obtained by the Dark Energy Survey (DES) \citep{Abbott18}.
We can then conclude that no modifications of the standard paradigm are expected as intermediate redshifts are involved.

Future efforts will be dedicated to the use of our new technique to fix refined constraints over dynamical dark energy models. Also, we will compare our outcomes with respect to previous model-independent calibrations.


\section*{Acknowledgements}
The authors are grateful to the anonymous referee for helpful suggestions which helped to improve the quality of the manuscript. R.D. and O.L. are grateful to Anna Silvia Baldi and Kuantay Boshkayev for useful discussions. L.A. acknowledges financial contribution from the agreement ASI-INAF n.2017-14-H.O. The work was supported in part by the Ministry of Education and Science of the Republic of Kazakhstan, Program 'Fundamental and applied studies in related fields of physics of terrestrial, near-earth and atmospheric processes and their practical application' IRN: BR05236494.


\bibliographystyle{mnras}
\bibliography{biblio}


\appendix

\begin{table*}
\centering
\scriptsize
\caption{List of the full sample of GRBs used in this work and their redshift $z$ and calibrated $\mu_{\rm GRB}$.}
\begin{tabular}{lcc|lcc|lcc|lcc}
\hline\hline
GRB & $z$ & $\mu_{\rm GRB}\pm\sigma_{\mu,{\rm GRB}}$ &  GRB & $z$ & $\mu_{\rm GRB}\pm\sigma_{\mu,{\rm GRB}}$ &  GRB & $z$ & $\mu_{\rm GRB}\pm\sigma_{\mu,{\rm GRB}}$ &  GRB & $z$ & $\mu_{\rm GRB}\pm\sigma_{\mu,{\rm GRB}}$ \\
\hline
970228	&	$	0.695	$	&	$	43.76	\pm	0.77	$	&	051109A	&	$	2.346	$	&	$	47.73	\pm	0.89	$	&	090323	&	$	3.57	$	&	$	47.08	\pm	0.53	$	&	120909A	&	$	3.93	$	&	$	48.20	\pm	0.93	$\\
970508	&	$	0.835	$	&	$	44.64	\pm	0.73	$	&	060115	&	$	3.5328	$	&	$	47.67	\pm	1.07	$	&	090328	&	$	0.736	$	&	$	45.47	\pm	0.36	$	&	121128A	&	$	2.2	$	&	$	45.56	\pm	0.30	$\\
970828	&	$	0.958	$	&	$	43.87	\pm	0.51	$	&	060124	&	$	2.296	$	&	$	46.47	\pm	0.88	$	&	090418A	&	$	1.608	$	&	$	48.05	\pm	0.64	$	&	130408A	&	$	3.758	$	&	$	48.55	\pm	0.46	$\\
971214	&	$	3.42	$	&	$	47.97	\pm	0.51	$	&	060206	&	$	4.0559	$	&	$	49.11	\pm	1.22	$	&	090423	&	$	8.1	$	&	$	50.05	\pm	0.67	$	&	130420A	&	$	1.297	$	&	$	42.87	\pm	0.29	$\\
980613	&	$	1.096	$	&	$	46.06	\pm	1.11	$	&	060210	&	$	3.91	$	&	$	47.52	\pm	0.79	$	&	090424	&	$	0.544	$	&	$	42.67	\pm	0.30	$	&	130427A	&	$	0.3399	$	&	$	41.59	\pm	0.41	$\\
980703	&	$	0.966	$	&	$	45.09	\pm	0.37	$	&	060218	&	$	0.03351	$	&	$	34.60	\pm	0.42	$	&	090516	&	$	4.109	$	&	$	47.93	\pm	1.00	$	&	130505A	&	$	2.27	$	&	$	46.27	\pm	0.39	$\\
981226	&	$	1.11	$	&	$	44.03	\pm	1.13	$	&	060306	&	$	3.5	$	&	$	47.48	\pm	1.04	$	&	090618	&	$	0.54	$	&	$	40.54	\pm	0.30	$	&	130518A	&	$	2.488	$	&	$	46.31	\pm	0.37	$\\
990123	&	$	1.6	$	&	$	45.37	\pm	0.70	$	&	060418	&	$	1.489	$	&	$	45.89	\pm	0.64	$	&	090715B	&	$	3.	$	&	$	47.01	\pm	0.78	$	&	130701A	&	$	1.155	$	&	$	44.69	\pm	0.30	$\\
990506	&	$	1.3	$	&	$	43.74	\pm	0.57	$	&	060526	&	$	3.22	$	&	$	45.97	\pm	0.50	$	&	090812	&	$	2.452	$	&	$	48.61	\pm	0.88	$	&	130831A	&	$	0.4791	$	&	$	41.24	\pm	0.30	$\\
990510	&	$	1.619	$	&	$	45.14	\pm	0.34	$	&	060607A	&	$	3.075	$	&	$	47.57	\pm	0.65	$	&	090902B	&	$	1.822	$	&	$	46.04	\pm	0.40	$	&	131011A	&	$	1.874	$	&	$	44.68	\pm	0.39	$\\
990705	&	$	0.842	$	&	$	43.51	\pm	0.73	$	&	060614	&	$	0.125	$	&	$	38.88	\pm	2.58	$	&	090926	&	$	2.1062	$	&	$	44.75	\pm	0.35	$	&	131030A	&	$	1.295	$	&	$	43.82	\pm	0.31	$\\
990712	&	$	0.434	$	&	$	41.85	\pm	0.44	$	&	060707	&	$	3.424	$	&	$	47.74	\pm	0.68	$	&	090926B	&	$	1.24	$	&	$	44.51	\pm	0.30	$	&	131105A	&	$	1.686	$	&	$	45.26	\pm	0.59	$\\
991208	&	$	0.706	$	&	$	41.98	\pm	0.31	$	&	060729	&	$	0.543	$	&	$	42.55	\pm	1.22	$	&	091003	&	$	0.8969	$	&	$	45.53	\pm	0.52	$	&	131108A	&	$	2.4	$	&	$	47.08	\pm	0.36	$\\
991216	&	$	1.02	$	&	$	43.36	\pm	0.52	$	&	060814	&	$	1.9229	$	&	$	45.63	\pm	0.79	$	&	091018	&	$	0.971	$	&	$	42.94	\pm	1.05	$	&	131117A	&	$	4.042	$	&	$	46.99	\pm	0.46	$\\
000131	&	$	4.5	$	&	$	47.18	\pm	1.04	$	&	060908	&	$	1.8836	$	&	$	47.14	\pm	1.18	$	&	091020	&	$	1.71	$	&	$	46.52	\pm	0.39	$	&	131231A	&	$	0.642	$	&	$	41.55	\pm	0.30	$\\
000210	&	$	0.846	$	&	$	44.82	\pm	0.34	$	&	060927	&	$	5.46	$	&	$	47.84	\pm	0.70	$	&	091024	&	$	1.092	$	&	$	43.92	\pm	0.32	$	&	140206A	&	$	2.73	$	&	$	46.07	\pm	0.33	$\\
000418	&	$	1.12	$	&	$	43.97	\pm	0.35	$	&	061007	&	$	1.262	$	&	$	44.36	\pm	0.41	$	&	091029	&	$	2.752	$	&	$	46.12	\pm	0.68	$	&	140213A	&	$	1.2076	$	&	$	43.72	\pm	0.30	$\\
000911	&	$	1.06	$	&	$	45.76	\pm	0.58	$	&	061121	&	$	1.314	$	&	$	46.73	\pm	0.40	$	&	091127	&	$	0.49	$	&	$	39.90	\pm	0.31	$	&	140419A	&	$	3.956	$	&	$	47.83	\pm	0.84	$\\
000926	&	$	2.07	$	&	$	44.62	\pm	0.37	$	&	061126	&	$	1.1588	$	&	$	46.15	\pm	0.76	$	&	091208B	&	$	1.063	$	&	$	45.18	\pm	0.30	$	&	140423A	&	$	3.26	$	&	$	47.24	\pm	0.40	$\\
010222	&	$	1.48	$	&	$	44.52	\pm	0.34	$	&	061222A	&	$	2.088	$	&	$	46.87	\pm	0.51	$	&	100414A	&	$	1.368	$	&	$	45.92	\pm	0.37	$	&	140506A	&	$	0.889	$	&	$	45.78	\pm	0.99	$\\
010921	&	$	0.45	$	&	$	42.29	\pm	0.49	$	&	070125	&	$	1.547	$	&	$	45.08	\pm	0.44	$	&	100621A	&	$	0.542	$	&	$	41.88	\pm	0.41	$	&	140508A	&	$	1.027	$	&	$	43.69	\pm	0.32	$\\
011121	&	$	0.36	$	&	$	44.03	\pm	0.70	$	&	070521	&	$	1.35	$	&	$	45.67	\pm	0.36	$	&	100728A	&	$	1.567	$	&	$	44.83	\pm	0.34	$	&	140512A	&	$	0.725	$	&	$	44.31	\pm	1.41	$\\
011211	&	$	2.14	$	&	$	45.34	\pm	0.35	$	&	071003	&	$	1.604	$	&	$	47.79	\pm	0.45	$	&	100728B	&	$	2.106	$	&	$	47.03	\pm	0.39	$	&	140515A	&	$	6.32	$	&	$	49.32	\pm	0.71	$\\
020124	&	$	3.198	$	&	$	46.59	\pm	0.79	$	&	071010B	&	$	0.947	$	&	$	42.48	\pm	0.59	$	&	100814A	&	$	1.44	$	&	$	43.93	\pm	0.36	$	&	140518A	&	$	4.707	$	&	$	47.58	\pm	0.46	$\\
020405	&	$	0.69	$	&	$	43.02	\pm	0.31	$	&	071020	&	$	2.145	$	&	$	48.45	\pm	0.66	$	&	100816A	&	$	0.8049	$	&	$	45.58	\pm	0.31	$	&	140620A	&	$	2.04	$	&	$	45.44	\pm	0.30	$\\
020813	&	$	1.25	$	&	$	43.73	\pm	0.67	$	&	071117	&	$	1.331	$	&	$	46.77	\pm	1.30	$	&	100906A	&	$	1.727	$	&	$	43.93	\pm	0.41	$	&	140623A	&	$	1.92	$	&	$	48.26	\pm	1.12	$\\
020819B	&	$	0.41	$	&	$	41.07	\pm	0.75	$	&	080207	&	$	2.0858	$	&	$	45.41	\pm	1.78	$	&	101213A	&	$	0.414	$	&	$	43.63	\pm	1.00	$	&	140629A	&	$	2.275	$	&	$	46.46	\pm	0.50	$\\
020903	&	$	0.25	$	&	$	39.31	\pm	1.38	$	&	080319B	&	$	0.937	$	&	$	44.01	\pm	0.36	$	&	101219B	&	$	0.55	$	&	$	42.89	\pm	0.31	$	&	140801A	&	$	1.32	$	&	$	44.79	\pm	0.30	$\\
021004	&	$	2.3	$	&	$	46.86	\pm	1.06	$	&	080411	&	$	1.03	$	&	$	44.50	\pm	0.38	$	&	110106B	&	$	0.618	$	&	$	44.31	\pm	0.68	$	&	140808A	&	$	3.29	$	&	$	48.53	\pm	0.45	$\\
021211	&	$	1.01	$	&	$	44.21	\pm	0.97	$	&	080413A	&	$	2.433	$	&	$	47.75	\pm	0.78	$	&	110205A	&	$	2.22	$	&	$	46.20	\pm	0.98	$	&	140907A	&	$	1.21	$	&	$	45.20	\pm	0.30	$\\
030226	&	$	1.98	$	&	$	45.23	\pm	0.55	$	&	080413B	&	$	1.1	$	&	$	44.63	\pm	0.70	$	&	110213A	&	$	1.46	$	&	$	44.71	\pm	0.79	$	&	141028A	&	$	2.33	$	&	$	46.12	\pm	0.35	$\\
030323	&	$	3.37	$	&	$	48.08	\pm	1.06	$	&	080603B	&	$	2.69	$	&	$	46.77	\pm	1.02	$	&	110213B	&	$	1.083	$	&	$	43.79	\pm	0.43	$	&	141109A	&	$	2.993	$	&	$	47.27	\pm	0.71	$\\
030328	&	$	1.52	$	&	$	43.58	\pm	0.43	$	&	080605	&	$	1.64	$	&	$	46.00	\pm	0.65	$	&	110422A	&	$	1.77	$	&	$	43.76	\pm	0.32	$	&	141220A	&	$	1.3195	$	&	$	43.78	\pm	0.33	$\\
030329	&	$	0.1685	$	&	$	38.28	\pm	0.30	$	&	080607	&	$	3.036	$	&	$	47.24	\pm	0.44	$	&	110503A	&	$	1.613	$	&	$	45.56	\pm	0.34	$	&	141221A	&	$	1.452	$	&	$	46.71	\pm	0.47	$\\
030429	&	$	2.65	$	&	$	46.09	\pm	0.50	$	&	080721	&	$	2.591	$	&	$	47.33	\pm	0.46	$	&	110715A	&	$	0.82	$	&	$	43.42	\pm	0.30	$	&	141225A	&	$	0.915	$	&	$	45.43	\pm	0.41	$\\
030528	&	$	0.78	$	&	$	41.06	\pm	0.41	$	&	080804	&	$	2.2045	$	&	$	47.83	\pm	0.35	$	&	110731A	&	$	2.83	$	&	$	47.76	\pm	0.37	$	&	150206A	&	$	2.087	$	&	$	45.71	\pm	0.44	$\\
040912B	&	$	1.563	$	&	$	42.91	\pm	2.16	$	&	080913	&	$	6.695	$	&	$	50.73	\pm	1.26	$	&	110801A	&	$	1.858	$	&	$	45.94	\pm	1.06	$	&	150301B	&	$	1.5169	$	&	$	46.73	\pm	0.52	$\\
040924	&	$	0.859	$	&	$	43.48	\pm	0.81	$	&	080916A	&	$	0.689	$	&	$	44.44	\pm	0.30	$	&	110818A	&	$	3.36	$	&	$	48.79	\pm	0.56	$	&	150314A	&	$	1.758	$	&	$	45.62	\pm	0.35	$\\
041006	&	$	0.716	$	&	$	41.64	\pm	0.57	$	&	080928	&	$	1.6919	$	&	$	43.64	\pm	0.62	$	&	111107A	&	$	2.893	$	&	$	48.38	\pm	0.72	$	&	150323A	&	$	0.593	$	&	$	44.32	\pm	0.40	$\\
041219	&	$	0.31	$	&	$	40.24	\pm	0.65	$	&	081007	&	$	0.5295	$	&	$	42.92	\pm	0.59	$	&	111228A	&	$	0.716	$	&	$	40.63	\pm	0.34	$	&	150403A	&	$	2.06	$	&	$	46.20	\pm	0.42	$\\
050318	&	$	1.4436	$	&	$	44.22	\pm	0.52	$	&	081008	&	$	1.9685	$	&	$	45.26	\pm	0.49	$	&	120119A	&	$	1.728	$	&	$	45.22	\pm	0.34	$	&	150413A	&	$	3.139	$	&	$	45.92	\pm	0.98	$\\
050401	&	$	2.8983	$	&	$	46.12	\pm	0.61	$	&	081028	&	$	3.038	$	&	$	45.51	\pm	0.95	$	&	120326A	&	$	1.798	$	&	$	45.04	\pm	0.30	$	&	150514A	&	$	0.807	$	&	$	43.47	\pm	0.43	$\\
050416A	&	$	0.6535	$	&	$	41.8	\pm	0.54	$	&	081118	&	$	2.58	$	&	$	45.21	\pm	0.30	$	&	120624B	&	$	2.1974	$	&	$	45.89	\pm	0.47	$	&	150821A	&	$	0.755	$	&	$	44.07	\pm	1.20	$\\
050525A	&	$	0.606	$	&	$	42.13	\pm	0.36	$	&	081121	&	$	2.512	$	&	$	46.48	\pm	0.34	$	&	120711A	&	$	1.405	$	&	$	46.00	\pm	0.40	$	&	151021A	&	$	1.49	$	&	$	43.82	\pm	0.36	$\\
050603	&	$	2.821	$	&	$	47.76	\pm	0.37	$	&	081203A	&	$	2.05	$	&	$	47.99	\pm	1.28	$	&	120712A	&	$	4.1745	$	&	$	48.38	\pm	0.53	$	&	151027A	&	$	0.81	$	&	$	44.57	\pm	1.31	$\\
050820	&	$	2.615	$	&	$	47.03	\pm	0.55	$	&	081221	&	$	2.26	$	&	$	44.55	\pm	0.31	$	&	120716A	&	$	2.486	$	&	$	45.60	\pm	0.32	$	&	151029A	&	$	1.423	$	&	$	45.24	\pm	0.51	$\\
050904	&	$	6.295	$	&	$	50.94	\pm	0.88	$	&	081222	&	$	2.77	$	&	$	47.00	\pm	0.34	$	&	120724A	&	$	1.48	$	&	$	44.25	\pm	0.68	$	&		&	$		$	&	$			$\\
050922C	&	$	2.199	$	&	$	47.20	\pm	0.73	$	&	090102	&	$	1.547	$	&	$	47.01	\pm	0.37	$	&	120802A	&	$	3.796	$	&	$	46.82	\pm	0.84	$	&		&	$		$	&	$				$\\
051022	&	$	0.809	$	&	$	43.30	\pm	0.82	$	&	090205	&	$	4.6497	$	&	$	49.78	\pm	0.88	$	&	120811C	&	$	2.671	$	&	$	45.95	\pm	0.30	$	&		&	$		$	&	$				$\\
\hline\hline
\end{tabular}
\label{tab:grbmuz}
\end{table*}


\bsp	
\label{lastpage}
\end{document}